\documentclass[preprint,prc,showpacs,preprintnumbers,
               superscriptaddress,floatfix]{revtex4}
\usepackage{graphicx,epsfig,latexsym,amssymb}
\usepackage{multirow,amsmath,array,booktabs}
\usepackage[section]{placeins}
\renewcommand{\dfrac}{\displaystyle\frac}
\date{\today}
\begin{document}

\title{Shell-model-Like APproach (SLAP) for the nuclear properties
in relativistic mean field theory}

\author{J. MENG}
\email{mengj@pku.edu.cn} \affiliation{School of Physics, Peking
University, Beijing 100871, P.R. China} \affiliation{Institute of
Theoretical Physics, Chinese Academy of Sciences, Beijing 100080,
P.R. China} \affiliation{Center of Theoretical Nuclear Physics,
National Laboratory of Heavy Ion Accelerator, Lanzhou 730000, P.R.
China}

\author{J. Y. GUO}
\affiliation{School of Physics, Peking University, Beijing 100871, P.R. China}
\affiliation{Department of Physics, Anhui University, Hefei 230039, P.R. China}

\author{L. LIU}
\affiliation{School of Physics, Peking University, Beijing 100871,
P.R. China}

\author{S. Q. ZHANG}
\affiliation{School of Physics, Peking University, Beijing 100871,
P.R. China}

\begin{abstract}

A Shell-model-Like APproach (SLAP) suggested to treat the pairing
correlations in relativistic mean field theory is introduced, in
which the occupancies thus obtained having been iterated back into
the densities. The formalism and numerical techniques are given in
detail. As examples, the ground state properties and low-lying
excited states for Ne isotopes are studied. The results thus
obtained are compared with the data available. The binding energies,
the odd-even staggering, as well as the tendency for the change of
the shapes in Ne isotopes are correctly reproduced.

\end{abstract}

 \maketitle

\section{Introduction}

The relativistic mean field (RMF) theory is one of the most
successful microscopic models \cite{Serot86}. During the past two
decades, it has received wide attention due to its success in
describing many nuclear phenomena for the stable nuclei
\cite{Reinhard89,Ring96} as well as nuclei even far from stability
\cite{meng98npa,meng05ppnp}. It has been shown that the relativistic
Brueckner theory can reproduce better the nuclear saturation
properties (the Coester line) in nuclear matter \cite{Brockmann90},
present a new explanation for the identical bands in superdeformed
nuclei \cite{konig93} and the neutron halo \cite{meng96prl}, predict
a new phenomenon --- giant neutron halos in heavy nuclei close to
the neutron drip line \cite{meng98prl}, give naturally the
spin-orbit potential, the origin of the pseudospin symmetry
\cite{Arima69,Hecht69} as a relativistic symmetry
\cite{Ginocchio97,meng98prc,meng99prc} and spin symmetry in the
anti-nucleon spectrum \cite{Zhou03prl}, and present good description
for the magnetic rotation \cite{Mad00} and the collective multipole
excitations \cite{Ma02}, etc.

Pairing correlations play an essential role in nuclear properties
such as, binding energies, odd-even effects, single particle orbit
occupation, electric-magnetic transition, low-lying collective
modes, moment of inertia, halo phenomena, etc. In order to be
realistic and to describe the open shell nuclei, the RMF theory must
be supplemented with the proper treatment of the pairing
correlations. This aspect becomes all the more important when
considering the continuum effects, which are crucial for the
description of drip line nuclei. So far the most commonly used
methods, the Bardeen-Cooper-Schrieffer (BCS) approximation and
Bogoliubov transformation, have become standard in the nuclear
physics literatures~\cite{Gambhir90,Geng03ptp}, even for exotic
nuclei where the Bogoliubov transformation is required to be done in
coordinate space~\cite{meng96prl,
meng98npa,Dobaczewski84,Dobaczewski96} or the introduction of the
resonant states is necessary~\cite{Yang01, Cao02, Sandulescu03,
Zhang04}.

As the approximate product of the quasi-particle wave functions in
quasi-particle formalism breaks the gauge symmetries connected
with the particle numbers, the projection methods are introduced
to restore the broken symmetries~\cite{PY.57}. There exists a vast
amount of literature on such projection methods (see, for
instance, Ref.~\cite{RS.80} and the references therein). The
variation after projection (VAP)~\cite{Zeh.67} is the appropriate
tool that fulfills the variation principle and provides a
self-consistent description of fluctuations going beyond
mean-field. Although, the method of variation after projection has
been known long time ago, the numerical solution of the
corresponding variation equations is relatively complicated. A
fully self-consistent variation after exact projection has been
carried out so far only in a limited number of
cases~\cite{RS.80,HS95}. Exact projection within full HFB-theory
is possible by a search of the minimum in the projected-energy
surface by gradient methods~\cite{ER.82a}. However these methods
are numerically very complicated and have been applied only to the
case of number-projection in restricted spaces and for separable
forces~\cite{ER.82b}. The problem whether simple HFB-like
equations can be obtained with the projected energy functional has
been recently addressed~\cite{SR.00} and it has been shown that
the variation of an arbitrary energy functional, which is
completely expressed in terms of the HFB density matrix $\rho$ and
the pairing tensor $\kappa$, results in the HFB like equations
with modified expressions for the pairing potential and the
Hartree-Fock field~\cite{SR.02}. This would allow the usage of
numerical algorithms for diagonalization of matrices as is done in
the ordinary HFB theory.

Meanwhile, the defects including the violation of the particle
number conservation and improbable treatment for the Pauli blocking
effects~\cite{Zeng83,Zeng941,Rowe70,Chasman90} in both the BCS
approximation and Bogoliubov transformation for finite fermion
system like nuclei can be avoided in the Shell-model-Like APproach
(SLAP)~\cite{Zeng83}, which was originally refereed as Particle
Number Conserving (PNC) method there.

Compared with the complicated particle number projection
technique, the SLAP proposed in the 1980s~\cite{Zeng83} avoids all
the difficulties encountered in BCS and Bogoliubov approximation
and takes into account the Pauli blocking effects strictly by
diagonalizing the pairing Hamiltonian directly in a reasonable
configuration space.  It has been found that the number of the
configurations with important weights in the low-lying solutions
is quite limited, only a few lowest-energy configurations
contribute significantly~\cite{Wu891prc}. This conclusion has been
used as the guideline of a many-body configuration truncation by
retaining a certain number of the lowest energy states. Extensive
studies and discussion on the validity of the truncated many-body
spaces as well as its application can be found in
Refs.~\cite{Wu891prc,Wu91prl,Zeng942,Zeng01prc,Zeng02prc} and
references therein. The application of the SLAP to solve the
nuclear mean-field plus pairing Hamiltonian problem with a
realistic deformed Woods-Saxon single-particle potential has been
reported in Ref.~\cite{Molique97}. The optimal basis construction
has been discussed and the stability of the final result with
respect to the basis cutoff has been illustrated. In particular
the presence of the low-lying seniority $s=0$ solutions, that are
usually poorly described in terms of the standard BCS or HFB
approximation, has been found to play a role in the interpretation
of the spectra of rotating nuclei.

As SLAP is based on the direct diagonalization of the pairing
Hamiltonian in the multi-particle configuration space, it has been
shown to be more accurate than the BCS calculation as compared
with the exact solution at least for the small
systems~\cite{Molique97}. Furthermore, the blocking effects are
taken into account automatically and both odd-A and even-even
nuclei can be treated on the same footing. In addition, the
excitation spectrum can be obtained conveniently in the present
algorithm.

In this paper, SLAP is suggested to treat the pairing correlations
in relativistic mean field theory, in which the occupancies thus
obtained having been iterated back into the densities. A brief
formalism is presented in section II. The numerical details and its
application for the ground state properties and low-lying excited
spectrum for Ne isotopes as well as the comparison with data are
given in section III.  Finally, a brief summary is given in section
IV.

\section{Formalism}

\subsection{Relativistic Mean Field Theory}

In the framework of RMF theory, the nuclear effective interaction
is usually described by the exchange of three mesons: the scalar
meson $\sigma$, which mediates the medium-range attraction between
the nucleons, the vector meson $\omega^{\mu}$, which mediates the
short-range repulsion, and the isovector-vector meson
$\vec{\rho}^{\mu}$, which provides the isospin dependence of the
nuclear force. The effective lagrangian density is as the
following:
\begin{eqnarray}
 {\cal L} &=&\bar{\psi}\left( i\gamma^{\mu }\partial_{\mu }-M\right)\psi
      +\dfrac{1}{2}\partial^{\mu }\sigma \partial_{\mu }\sigma
      -\dfrac{1}{2}m_{\sigma}^{2}\sigma^{2}-\dfrac{1}{3}g_{2}\sigma^{3}
      -\dfrac{1}{4}g_{3}\sigma^{4}
      -g_{\sigma}\bar{\psi}\sigma \psi \notag \\
    &&-\dfrac{1}{4}\Omega^{\mu \nu }\Omega_{\mu \nu}
      +\dfrac{1}{2}m_{\omega}^{2}\omega^{\mu }\omega_{\mu}
      -g_{\omega }\bar{\psi}\gamma^{\mu }\psi\omega_{\mu}
      +\dfrac{1}{4}g_4(\omega^\mu\omega_\mu)^2  \notag \\
    &&-\dfrac{1}{4}\vec{R}^{\mu \nu}\vec{R}_{\mu \nu}
      +\dfrac{1}{2}m_{\rho}^{2}\vec{\rho}^{\mu}\vec{\rho}_{\mu }
      -g_{\rho}\bar{\psi}\gamma^{\mu}\vec{\tau}\psi \vec{\rho}_{\mu} \notag \\
    &&-\dfrac{1}{4}F^{\mu \nu }F_{\mu \nu }
     -e\bar{\psi}\gamma^{\mu}\dfrac{1-\tau _{3}}{2}A_{\mu }\psi,
\end{eqnarray}%
where the field tensors of the vector mesons and the
electromagnetic field take the forms:
\begin{equation}
\left\{
\begin{array}{lll}
    \Omega^{\mu \nu } & = & \partial^{\mu }\omega^{\nu }
                             -\partial^{\nu}\omega^{\mu }, \\
    \vec{R}^{\mu \nu }  & = & \partial^{\mu }\vec{\rho}^{\nu }
                            -\partial^{\nu }\vec{\rho}^{\mu}
                            -2g_{\rho }\vec{\rho}^{\mu}\times\vec{\rho}^{\nu }, \\
    F^{\mu \nu } & = & \partial^{\mu }A^{\nu }-\partial^{\nu }A^{\mu },
\end{array}
\right.
\end{equation}
and other symbols have their usual meanings
\cite{Serot86,Reinhard89,Ring96}.

The classical variation principle leads to the Dirac equation,
\begin{equation}
      \left\{ -i\mathbf{\alpha}\cdot\mathbf{\nabla}
       + V(\mathbf{r})+\beta \left[
       M+S(\mathbf{r})\right]
      \right\} \psi_{i}=\epsilon_{i}\psi_{i},
\end{equation}
for the nucleon spinors and the Klein-Gordon equations,
\begin{equation}
\left\{
\begin{array}{rcl}
      \left[ -\Delta +m_{\sigma}^{2}\right]\sigma(\mathbf{r})
             & = & -g_{\sigma}\rho _{s}(\mathbf{r})
                   -g_{2}\sigma^{2}(\mathbf{r})
                   -g_{3}\sigma^{3}(\mathbf{r}), \\
      \left[ -\Delta +m_{\omega}^{2}\right] \omega^{\mu }(\mathbf{r})
             & = &  g_{\omega }j^{\mu }(\mathbf{r})
                   -g_4(\omega^\nu\omega_\nu)\omega^\mu(\mathbf{r}), \\
      \left[ -\Delta +m_{\rho }^{2}\right] \vec{\rho}^{\mu }(\mathbf{r})
             & = &  g_{\rho}\vec{j}^{\mu }(\mathbf{r}), \\
      -\Delta A^{\mu }(\mathbf{r})
             & = & ej_{p}^{\mu }(\mathbf{{r}),}%
\end{array}%
\right.
\end{equation}%
for the mesons, where
\begin{equation}
\left\{
\begin{array}{lll}
       V(\mathbf{r})
            & = & \beta\left[g_{\omega}\gamma^\mu\omega_{\mu}(\mathbf{r})
                 +g_{\rho}\gamma^\mu\vec{\tau}\vec{\rho}_{\mu}(\mathbf{r})
                 +e\gamma^\mu\dfrac{1-\tau_{3}}{2}A_{\mu}(\mathbf{r})\right], \\
       S(\mathbf{r}) & = & g_{\sigma }\sigma(\mathbf{r}),%
\end{array}
\right.
\end{equation}
are respectively the vector and scalar potentials and the source
terms for the mesons and the photons are
\begin{equation}
\left\{
\begin{array}{lll}
      \rho _{s}(\mathbf{r}) & = & \sum \limits_{i=1}^{A}\bar{\psi _{i}}\psi _{i}, \\
      j^{\mu }(\mathbf{r})  & = & \sum \limits_{i=1}^{A}\bar{\psi_{i}}\gamma^{\mu }\psi_{i}, \\
      \vec{j}^{\mu }(\mathbf{r}) & = & \sum \limits_{i=1}^{A}\bar{\psi_{i}}\gamma ^{\mu }\vec{\tau}\psi _{i}, \\
      j_{p}^{\mu}(\mathbf{r})& = &\sum \limits_{i=1}^{A}\bar{\psi_{i}}\gamma^{\mu}\dfrac{1-\tau_{3}}{2}\psi _{i}.%
\end{array}%
\right.
\end{equation}%
In Eqs.(6), the summations are taken over the valence nucleons only,
i.e., no-sea approximation is adopted. The coupled equations (3) and
(4) are non-linear quantum field equations, and their solutions are
very complicated. Therefore, the mean field approximation is
generally used to solve Eqs.(3) and (4): i.e., the meson field
operators in Eq.(4) are replaced by their corresponding expectation
values, and the nucleons are considered to move independently in the
classical meson fields. In such a way the coupled equations (3) and
(4) can be self-consistently solved by iteration.

The symmetries of the system simplify the calculations
considerably. In the system considered in this work, as there
exists time reversal symmetry, there are no currents in the
nucleus and the spatial vector components of $\mathbf{\omega}$,
$\vec{\mathbf{\rho}}$ and $\mathbf{A}$ vanish. This
leaves only the time-like components $\omega^{0}$, $\vec{\rho}^{0}$ and $A^{0}$%
. Charge conservation guarantees that only the 3-component of the isovector $%
\vec{\rho}$ survives. For simplicity, we denote it in the
following simply by $\rho^0$.

For axially deformed nuclei, i.e., the systems which have
rotational symmetry around a symmetrical axis (assumed to be the
$z$-axis), it is useful to work with cylindrical coordinates:
$x=r_{\perp }\cos \varphi ,y=r_{\perp }\sin \varphi $ and $z$. The
potential of the nucleon and the sources of meson fields depend
only on the coordinates $r_{\perp }$ and $z$.
The Dirac spinor $\psi _{i}$ for the nucleon with the index $i$
is characterized by the quantum numbers $%
\Omega _{i}$, $\pi _{i}$ and $t_{i}$, where $\Omega
_{i}=m_{l_{i}}+m_{s_{i}}$ is the eigenvalue of the angular momentum
operator $J_{z}$, $\pi _{i}$ is the parity and $t_{i}$ is the
isospin, i.e., the Dirac spinor $\psi_i$ has the
form~\cite{Gambhir90},
\begin{equation}
\psi _{i}=\left(
\begin{array}{c}
      f_{i}(\mathbf{r}) \\
      ig_{i}(\mathbf{r})%
\end{array}%
\right) =\dfrac{1}{\sqrt{2\pi }}\left(
\begin{array}{c}
      f_{i}^{+}(z,r_{\perp })e^{i(\Omega _{i}-1/2)\varphi } \\
      f_{i}^{-}(z,r_{\perp })e^{i(\Omega _{i}+1/2)\varphi } \\
      ig_{i}^{+}(z,r_{\perp })e^{i(\Omega _{i}-1/2)\varphi } \\
      ig_{i}^{-}(z,r_{\perp })e^{i(\Omega _{i}+1/2)\varphi }%
\end{array}%
\right) \chi _{t_{i}}(t).
\end{equation}%
For each solution with positive $\Omega _{i}$, $\psi _{i}$, there
is the time-reversed solution with the same energy, $\psi
_{\bar{i}}=T\psi _{i}$, in which the time reversal operator
$T=-i\sigma _{y}K$ and $K$ is the complex conjugation. For nuclei
with time reversal symmetry, the contributions to the densities of
the time reversal states, $i$ and $\bar{i}$, are identical.
Therefore, the densities can be represented as
\begin{equation}
       \rho _{s,v}=2\sum_{i>0}n_{i}\left[ \left( \left\vert f_{i}^{+}\right\vert^{2}
                   +\left\vert f_{i}^{-}\right\vert ^{2}\right)
                   \mp \left( \left\vert g_{i}^{+}\right\vert ^{2}
                   +\left\vert g_{i}^{-}\right\vert ^{2}\right) \right]
\end{equation}%
and, in a similar way, $\rho _{3}$ and $\rho _{c}$. The sum here
runs over only states with positive $\Omega _{i}$. The occupation
number for state $i$ is represented by $n_{i}$. These densities
serve as the sources for the fields $\phi =\sigma ,\omega
^{0},\rho ^{0}$ and $A^{0}$, which are determined by the
Klein-Gordon equation in cylindrical coordinates.

To solve the equations (3) and (4), the spinors $f_{i}^{\pm
}(z,r_{\perp })$ and $g_{i}^{\pm }(z,r_{\perp })$ can be expanded
in terms of the eigenfunctions of a deformed axially symmetric
oscillator potential \cite{Gambhir90} or Woods-Saxon
potential\cite{Zhou03prc}, and the solution of the problem is
transformed into a diagonalization of a Hermitian matrix. The
details can be found in Ref.\cite{Gambhir90,Zhou03prc}.

The total energy of the system is:
\begin{equation}
      \label{RMFE}
      E_{\text{RMF}}=E_{\text{nucleon}}+E_{\sigma }+E_{\omega }
                     +E_{\rho }+E_{c}+E_{\text{CM}}
\end{equation}%
with
\begin{equation*}
\left\{
\begin{array}{cll}
      E_{\text{nucleon}} & = & \sum \limits_{i}\epsilon _{i} \; ,  \\
      E_{\sigma } & = & -\dfrac{1}{2}\int d^{3}r\left\{ g_{\sigma }
                      \rho _{s}(\mathbf{r})\sigma (\mathbf{r})
                      +\left[ \dfrac{1}{3}g_{2}\sigma (\mathbf{r})^{3}
                      +\dfrac{1}{2}g_{3}\sigma (\mathbf{r})^{4}\right] \right\} , \\
      E_{\omega } & = & -\dfrac{1}{2}\int d^{3}r\left\{g_{\omega }
                      \rho _{v}(\mathbf{r})\omega^{0}(\mathbf{r})
                      -\dfrac{1}{2}g_4\omega^0(\mathbf{r})^4\right\}, \\
      E_{\rho } & = & -\dfrac{1}{2}\int d^{3}rg_{\rho }
                     \rho_{3}(\mathbf{r})\rho^{00}(\mathbf{r}), \\
      E_{c}   & =  & -\dfrac{e^{2}}{8\pi }\int d^{3}r\rho_{c}(\mathbf{r})A^{0}(\mathbf{r}), \\
      E_{\text{CM}} & = & -\dfrac{3}{4}41A^{-1/3},%
\end{array}%
\right.
\end{equation*}%
where $E_{\text{nucleon}}$ is the sum of the energy for nucleon
$\epsilon _{i}$, $E_{\sigma },E_{\omega },E_{\rho },$ and $%
E_{c} $ are the contributions of the meson fields and the Coulomb field, and $E_{%
\text{CM}}$ is the center of mass correction.

\subsection{SLAP for the Pairing Correlations}

Based on the single-particle levels and wavefunctions obtained from
the RMF theory, the SLAP can be adopted to treat the nuclear pairing
correlations, here after refereed as RMF+SLAP. The Hamiltonian
reads:
\begin{equation}
      H=H_{\text{s.p.}}+H_{\text{pair}},
\end{equation}
where, $H_{\text{s.p.}}=\sum \limits_{\nu }\epsilon _{\nu }a_{\nu
}^{+}a_{\nu }$ and $H_{\text{pair}}=-G\sum \limits_{\mu,\nu
>0}^{\mu\neq\nu}a_{\mu}^{+}a_{\bar{\mu}}^{+}a_{\bar{%
\nu}}a_{\nu }\text{,}$ with $\epsilon _{\nu}$
the single particle energy obtained from the RMF theory and $H_{%
\text{pair}}$ the pairing Hamiltonian with average strength $G$
simplified from the quantization of the meson fields, $\nu $ is
the notation of the level with the quantum numbers $\left(\Omega
,\pi ,t\right) $ for axially deformed nuclei, and $\bar{\nu}$ is
the time-reversal state of $\nu $.

The multi-particle configurations (MPC) are constructed as bases
to diagonalize the Hamiltonian in Eq.(10). For a even $N=2n$
particle system, the MPC are constructed as the following:

(a) The fully paired configuration (seniority $s=0$):
\begin{equation}
      |\rho _{1}\bar{\rho _{1}}\cdots \rho _{n}\bar{\rho _{n}}\rangle
      =a_{\rho_{1}}^{+}a_{\bar{\rho}_{1}}^{+}\cdots a_{\rho _{n}}^{+}
       a_{\bar{\rho}_{n}}^{+}\left\vert 0\right\rangle ,
\end{equation}

(b) The configuration with two unpaired particles (seniority
$s=2$):
\begin{equation}
      |\mu\bar{\nu}\rho _{1}\bar{\rho _{1}}\cdots
      \rho_{n-1}\bar{\rho}_{n-1}\rangle
         =a_{\mu }^{+}a_{\bar{\nu}}^{+}a_{\rho_{1}}^{+}
          a_{\bar{\rho}_{1}}^{+}\cdots
          a_{\rho _{n-1}}^{+}
          a_{\bar{\rho}_{n-1}}^{+}\left\vert 0\right\rangle ,
\end{equation}%
and so on~\cite{Zeng83}. The MPC for a odd $N=2n+1$ particle
system can be constructed similarly. For the axially deformed
nuclei, as the parity $\pi$, seniority $s$, and eigenvalue $K$ of
the third component for the total angular momentum operator
$J_{z}$ are good quantum numbers, the MPC space of a even particle
system can be written as the following:%
\begin{eqnarray}
      \text{MPC space} &=&\left( s=0,K=0^{+}\right)  \notag \\
                       &&\oplus \left( s=2,K=0^{+}\right)
                         \oplus \left( s=2,K=1^{+}\right)
                         \oplus \left( s=2,K=2^{+}\right)
                         \oplus \cdots  \notag \\
                       &&\oplus \left( s=4,K=0^{+}\right)
                         \oplus \left( s=4,K=1^{+}\right)
                         \oplus \left( s=4,K=2^{+}\right)
                         \oplus \cdots  \notag \\
                       &&\oplus \cdots .
\end{eqnarray}%
In realistic calculation, the MPC space has to be truncated and
 an energy cutoff $E_{c}$ is introduced. Only the configurations with
 energies $E_{i}-E_{0}\leq E_{c}$ are
used to diagonalize the Hamiltonian $H$ in Eq.(10), where $E_{i}$
and $E_{0}$ are the energies of the $i$th configuration and the
lowest configuration, respectively. The cutoff in the single
particle states is implicitly included in the energy cutoff $E_c$.

The corresponding nuclear wavefunction can be expanded as
\begin{eqnarray}
      \psi^{\beta } &=&\sum_{\rho _{1},\cdots ,
                       \rho _{n}}V_{\rho_{1},\cdots,
                       \rho _{n}}^{\beta }|\rho _{1}\bar{\rho _{1}}\cdots
                       \rho _{n}\bar{\rho _{n}}\rangle
                       +\sum_{\mu ,\nu }\sum_{\rho _{1},\cdots ,
                       \rho_{n-1}}V_{\rho _{1},\cdots ,\rho _{n-1}}^{\beta (\mu \nu )}
                       |\mu\bar{\nu}\rho _{1}\bar{\rho _{1}}\cdots
                       \rho_{n-1}\bar{\rho}_{n-1}\rangle +\cdots,
\end{eqnarray}
where $\beta =0$ (ground state), 1, 2, 3, $\cdots$ (excited
states).

The occupation probability of the $i$th-level for state $\beta$ is
\begin{equation}
      n_{i}^{\beta}=\sum_{\rho _{1},\cdots ,\rho _{n-1}}
                \left\vert V_{\rho_{1},\cdots ,\rho _{n-1},i}^{\beta}\right\vert ^{2}
                +\sum_{\mu ,\nu }\sum_{\rho _{1},\cdots,
                       \rho_{n-2}}\left\vert V_{\rho _{1},\cdots ,\rho _{n-2},i}
                       ^{\beta (\mu \nu )}\right\vert ^{2}+\cdots,
                \text{ \ \ \ }
                i=1, 2, 3, \cdots .
\end{equation}%
Replacing the occupation number $n_{i}$ in Eq.(8) by
$n_{i}^{\beta=0}$ , the sources in Eqs.(4) are obtained and the
occupancies thus obtained having been iterated back into the
densities, which give the new meson fields and new electromagnetic
field. These new fields then are used to calculate the vector and
scalar potentials in Eqs.(5). With these new potentials, the Dirac
equation Eq.(3) is solved again. These processes should be
repeated until the results converge to the given precision. As the
pairing correlations are taken into account,
the pairing energy $E_{\text{pair}}=\left\langle \psi ^{\beta}|H_{\text{pair}%
}|\psi ^{\beta}\right\rangle $ should be added to the total energy
in Eq.~(\ref{RMFE}).

\section{Results and discussions}

We have performed the RMF+SLAP calculations for the Ne isotopes with
the usual effective interactions, e.g.,  PK1\cite{Long04}, NLSH
\cite{Sharma93}, NL3 \cite{Lalazissis97}, and TMA\cite{Sugahara95}.
As the conclusions do not depend on the effective interactions used
here, only the results with the effective interaction PK1 are
presented in detail. Similar investigation has been done with the
RMF+BCS approach in Ref. \cite{Geng04npa}. For the axially deformed
nuclei, the parity $\pi$ and the eigenvalue $K$ of the third
component of total angular momentum operator $J_{z}$ are good
quantum numbers. In the numerical calculations, the Dirac equation
Eq.(3) for the nucleons and the Klein-Gordon equations Eqs.(4) for
the mesons and the photon are solved by expansion in the harmonic
oscillator basis with 14 oscillator shells for both the fermion
fields and the boson fields. The oscillator frequency of the
harmonic oscillator basis is fixed as $\hslash \omega
_{0}=41A^{-1/3}$ MeV and the deformation of harmonic oscillator
basis $\beta_0$ is reasonably set to obtain the lowest energy.

Taking $^{24}$Ne as an example, the detailed procedures of SLAP
for the pairing correlations are:

\begin{itemize}
  \item From the solution of the Dirac equation, the neutron
         and proton single-particle levels are obtained, as
         schematically shown in Fig. 1, where the dashed line
         represents the last occupied level in the filling
         approximation.
  \item With the single-particle levels thus obtained, the
         MPC are constructed according to the Eqs.(11) and (12) ( see also
         Ref. \cite{Zeng83} ) with given truncation energy $E_c$. The
         truncation energy $E_c$ is chosen in such a way that no configurations
         with important weights are omitted in the SLAP
         calculations. In the following calculations on the Ne isotopes,
         the truncation energy is reasonably fixed to $E_c=40$
         MeV.
         With this truncation energy, all the configurations with weight above
         $3.7\times10^{-6}$ have been included for $^{24}$Ne, while configurations
         with weight above 0.032\% included for $^{18}$Ne. More detailed discussion
         on the configuration truncation can be found in Refs.
         \cite{Zeng83,Zeng941,Molique97,Wu891prc,Wu91prl,Zeng942,Zeng01prc,Zeng02prc} and
         the references therein.
  \item As the parity $\pi$, seniority $s$, and $K$
         are good quantum numbers, the Hamiltonian $H$ in Eq.(10)
         is block diagonal in the MPC space Eq.(13). In every subspace, the single
         particle Hamiltonian $H_{\textrm{s.p.}}$ is diagonal.
         The non-zero off-diagonal matrix elements come from $H_{\textrm{pair}}$
         for configurations with a difference of
         one pair particles in the particle-pair occupation orbital.
         The pairing strength $G_\tau$=$C_\tau/A$ MeV ($\tau=p, n$) is adopted for all the Ne isotopes,
         as normally done in RMF+BCS calculations, e.g., see Ref.\cite{DelEstal01}.
         Here for truncation energy $E_c=40$ MeV, $C_\tau=10$ is fixed by reproducing the experimental
         odd-even mass difference of $^{24-26}$Ne.
  \item Diagonalizing the matrix of the Hamiltonian $H$ in Eq.(10) obtained above,
         the solutions for the ground state and the low-lying excited states are obtained.
         The occupation probabilities $n_{i}^{\beta}$ of the single-particle level ${i}$
         in state ${\beta}$ can be calculated from Eq.(15).
         Replacing the occupation number $n_{i}$ by $n_{i}^{\beta =0}$
         in Eq.(8), the densities in Eqs.(6) can be obtained, which give the new meson fields and new
         electromagnetic field. These new fields can be used to calculate the vector and scalar
         potentials in Eqs.(5). With these new vector and scalar potentials, the Dirac equation
         Eq.(3) is solved again.
  \item  The above processes should be repeated until the results converge to the given
         precision. The calculated results obtained in such a way
         will be refereed as RMF+SLAP in the following.

\end{itemize}

 As the pairing correlations are taken into account, for state $\beta$,
  the energy for nucleons $E_{\textrm{nucleon}}$ in
  the total energy of system Eq.(9) should be modified as:
         $$E_{\text{nucleon}}=\sum_{i}n_{i}^{\beta}\epsilon _{i},$$
  and the pairing energy
         $E_{\text{pair}}=\left\langle \psi ^{\beta}|H_{\text{pair}%
             }|\psi ^{\beta}\right\rangle $
  should be added to the total energy.

The results for the ground state of $^{24}$Ne from the
RMF+SLAP$^{*}$ and RMF+SLAP calculations with PK1 are shown in Table
I, including the binding energy $E$, binding energy per nucleon
$E/A$, the rms radii for neutron, proton, and matter, $R_{n}$,
$R_{p}$, and $ R_{m}$, and the quadruple deformation for neutron,
proton and matter, $\beta_{2n}$, $\beta_{2p}$ and $\beta_{2m}$. The
difference between these two calculations lies in that in
RMF+SLAP$^{*}$, the occupancies obtained by SLAP is not iterated
back into the densities. While in RMF+SLAP, the occupancies obtained
by SLAP is self-consistently iterated back into the densities. The
corresponding calculations and the data available
\cite{Audi95,Raman01} are included for comparisons. From Table I, it
can be seen that as pairing correlations are taken into account,
both the RMF+SLAP$^{*}$ and RMF+SLAP calculations give more bound
for the total binding energy and larger rms radii due to the
contribution from those loosely bound levels, which is similar as
that obtained from the Bogoliubov transformation, i.e., the pairing
correlations can increase the nuclear rms radius\cite{meng96prl}.
However, compared with RMF+SLAP$^{*}$, the binding energy from
RMF+SLAP calculations is less. Furthermore, the deformation of
nucleus reduced by the pairing correlations is realized only in
RMF+SLAP, which clearly demonstrated the necessity to iterate the
occupancies obtained by SLAP self-consistently back into the
densities.

As the Pauli blocking effects can be strictly treated in SLAP, a
self-consistent description for even-even, odd-A, and odd-odd nuclei
can be obtained in the RMF+SLAP calculations. In Fig.2, the
occupation probabilities for the neutron levels are displayed for
$^{24}$Ne in the left panel and for $^{25}$Ne in the right panel,
where the occupation probabilities of levels above [202]5/2$^+$ for
$^{24}$Ne and those of levels above [200]1/2$^+$ for $^{25}$Ne are
multiplied by a factor of 10 for clarity. For $^{24}$Ne, if without
pairing correlation, the levels above [202]5/2$^+$ will be
unoccupied and the others are fully occupied. In the RMF+SLAP
calculations, the levels above [202]5/2$^+$ have been occupied
partly, while the occupation probabilities for levels below Fermi
surface are decreased accordingly. For $^{25}$Ne, the [200]1/2$^+$
is the blocking level and is occupied by a single nucleon in the
RMF+SLAP calculations. The occupations for the levels beyond
[200]1/2$^+$ have been reduced due to the blocking effects. For
$^{24}$Ne,  both RMF+SLAP$^{*}$ and RMF+SLAP calculations have been
done and it turns out that their difference in the occupation
probabilities can be neglected.

The ground state properties of Ne isotopes including the odd-A and
even-even nuclei from the RMF+SLAP calculations with PK1 are listed
in Table II, including the total binding energy $E$, the binding
energy per nucleon $E/A$, the rms radii for neutron, proton, and
matter, $R_{n}$, $R_{p}$, and $ R_{m}$, and the quadruple
deformation for neutron, proton and matter, $\beta_{2n}$,
$\beta_{2p}$ and $\beta_{2m}$. The binding energies per nucleon for
Ne isotopes from the RMF+SLAP calculations are shown in Fig.3, in
comparison with the available data~\cite{Audi95}. It can seen that
for the ground state, the RMF+SLAP calculations, similar as the
RMF+BCS calculations~\cite{Geng04npa}, give very good description of
the data, except for the proton-rich nucleus $^{18}$Ne. However,
RMF+SLAP approximation can do more than RMF+BCS calculations, as
will be seen in the following.

 One- and two-neutron separation energies defined as,
\begin{equation}
    S_n(Z,N)=E(Z,N)-E(Z,N-1),
\end{equation}
and
\begin{equation}
   S_{2n}(Z,N)=E(Z,N)-E(Z,N-2),
\end{equation}
are sensitive quantities to test a microscopic theory, where
$E(Z,N)$ is the binding energy of nucleus with proton number $Z$
and neutron number $N$.

In Fig. 4, the one- and two-neutron separation energies from the
RMF+SLAP calculations (filled circles) with PK1 in comparison with
the data (open circles)\cite{Audi95}. From the figure, it is shown
that the experimental odd-even staggering is well reproduced in  the
RMF+SLAP calculations for either the one-neutron separation energy
or the two-neutron separation energy.

In Fig. 5, the deformations $\beta_{2}$ for Ne isotopes from the
RMF+SLAP calculations are given, including the data~\cite{Raman01}.
It should be mentioned that as the prolate or oblate shapes can not
be distinguished experimentally, the assumption of shape oblate for
$^{24}$Ne has been done. First of all, the tendency for the shape
change with neutron number in Ne isotopes is correctly reproduced.
The RMF+SLAP calculations give even better results than the RMF+BCS
calculations~\cite{Geng04npa}. From the figure, it can be seen that
the deformations are generally underestimated, particularly for the
proton rich side. In Ref.~\cite{Raman01}, the deformation parameter
$\beta$ is determined from $B(E2)\uparrow$ for the ground state to
$2^+$ state
\begin{equation}
   \beta=(4\pi/3ZR_0^2)[B(E2)\uparrow/e^2]^{1/2}, \label{eq:beta}
\end{equation}
where the radius $R_0$ has been taken to be $1.2 A^{1/3}$ fm.
However, it has been well demonstrated that the experiential
relation of radius is only suitable for medium-mass and heavy
nuclei~\cite{Zhang02}. While for light nuclei the estimation for
the radius $R_0$ is too small, this leads to a large deformation
$\beta$ with Eq.(\ref{eq:beta}). With this factor considered, the
consistency between the calculation results and the deformation
data will be more agreeable.

The neutron, proton and total pairing energies from the RMF+SLAP
calculations are displayed in Fig.6 as the open circles, open
squares and filled circles respectively. It can be seen that the
odd-even staggering has been clearly revealed for the neutron
pairing energy due to the correct treatment of the Pauli blocking
effects, while the proton pairing energy varies smoothly with the
neutron number $N$. The large neutron pairing energy at $N=20$ in
present calculations indicates disappearances of the traditional
magic number at $N=20$ in the Ne isotope chain. Although from Fig.5
the nucleus $^{30}$Ne is nearly spherical ($\beta_2 \sim 0.1$), the
magic number $N=20$ disappears. This indicates that the
disappearance of the magic number here is not due to the
deformation, instead it might come from the fact that the neutron
Fermi surface in $^{30}$Ne is very close to the threshold of
continuum, as noted and discussed in
Refs.\cite{meng98npa,mengplb98,mengprc02}.

In order to justify the description for nuclei close to the drip
line, the analytical continuation in coupling constant (ACCC)
method or the S-matrix method~\cite{Yang01,Sandulescu03,Zhang04}
should be adopted to resort the resonance states. The works along
this line is in progress.

The unified description for both the ground and excited states is
another advantage of the RMF+SLAP. By diagonalization of the
pairing Hamiltonian in the corresponding subspace with fixed $K$,
parity $\pi$, and seniority $s$, the spectra of multi-particle
collective excitations can be obtained.

As an example, the collective excitation spectra for $^{24}$Ne are
shown in Fig.8, where the subfigures (a), (b), (c) and (d)
respectively represent the spectra of positive and negative parity
for neutron, and those for proton. The calculated $0^{+}$ excitation
spectra include $E_{n}^{0_1^+}$=5.622 MeV for neutron and
$E_{p}^{0_1^+}$=4.912 MeV for proton, which are very close to the
only known experimental data $E_{exp}^{0_1^+}$=4.764 MeV.  From Fig.
8, it can also be seen that the RMF+SLAP can also be used to
investigate the excited states with high-$K$ value. It may provide
another self-consistent microscopic description and prediction of
the interesting high-$K$ energy trap\cite{Baldwin97,Walker99}.

\section{summary}

The Shell-model-Like APproach (SLAP) has been applied to treat the
pairing correlations in the framework of RMF theory, which can
exactly treat the Pauli blocking effects. The formalism and
numerical techniques are presented in detail and the Ne isotopes
are taken as an example to demonstrate the applicability of the
method.

First of all, the RMF+SLAP with effective interactions PK1 has been
used to describe the ground state properties of the Ne isotopes,
including the binding energies, one-neutron and two-neutron
separation energies, deformation, pairing energy and occupation
probability of single particle orbital, etc.. The calculated binding
energies from the RMF+SLAP agree well with the data and are
comparable with the RMF+BCS calculations. The odd-even staggering is
well reproduced, as shown in the neutron pairing energy and
one-neutron or two-neutron separation energy. The tendency for the
change of the shapes with neutron number in Ne isotopes is correctly
reproduced. Furthermore the RMF+SLAP calculations give better
results than the RMF+BCS calculations for $^{26}$Ne and $^{28}$Ne.

The other advantage of the RMF+SLAP method is the description of
both the ground and excited states simultaneously. As examples,
the collective excitation spectra for $^{24}$Ne with positive and
negative parity for both proton and neutron are presented. Its
potential application for the interesting high-$K$ energy trap has
been demonstrated.

In conclusion, SLAP is a good method to treat the pairing
correlations in the RMF theory, which can provide a good
description for the properties in not only the ground state but
also the excited states.

\begin{acknowledgments}
This work is partly supported by the Major State Basic Research
Development Program Under Contract Number G2000077407, the
National Natural Science Foundation of China under Grant No.
10025522, 10435010 and 10221003, the Doctoral Program Foundation
from the Ministry of Education in China, and China Postdoctoral
Science Foundation.
\end{acknowledgments}

\newpage

\begin{table}[!h]
\caption{The binding energy, radii and deformation of ground state
in $^{24}$Ne obtained from the RMF+SLAP$^{*}$ and RMF+SLAP
calculations with PK1, in comparison with the RMF calculation and
data available. Listed are total binding energy $E$, binding energy
per nucleon $E/A$, neutron, proton and matter root mean square
radii, $R_n$, $R_p$ and $R_m$, and the quadruple deformation
parameters for the neutron, proton and matter distributions,
$\beta_{2n}$, $\beta_{2p}$ and $\beta_{2m}$, respectively.}
\label{tab:Ne24ground}
\begin{center}
\begin{tabular}{c |c c|c c c|c c c}
    \hline\hline
      & E & E/A & $R_{n}$ & $R_{p}$ & $R_{m}$ & $\beta_{2n}$ & $\beta_{2p}$ & $\beta_{2m}$ \\
    \hline
    EXP            & -191.836 & -7.993 &       &       &       &        &        & -0.45 \\
    RMF            & -189.283 & -7.887 & 2.985 & 2.756 & 2.892 & -0.278 & -0.238 & -0.261 \\
    RMF+SLAP$^{*}$ & -190.723 & -7.947 & 2.994 & 2.757 & 2.898 & -0.278 & -0.236 & -0.261 \\
    RMF+SLAP       & -189.974 & -7.916 & 2.994 & 2.758 & 2.898 & -0.275 & -0.235 & -0.258 \\
    \hline
\end{tabular}%
\end{center}
\end{table}

\begin{table}[!h]
\caption{The ground state properties of Ne isotopes obtained from
the RMF+SLAP calculations with PK1, including the binding energy
$E$, binding energy per nucleon $E/A$, neutron, proton and matter
root mean square radii, $R_n$, $R_p$ and $R_m$, and the quadruple
deformation parameters for the neutron, proton and matter
distributions, $\beta_{2n}$, $\beta_{2p}$ and $\beta_{2m}$,
respectively, with $A$ the mass number and $N$ the neutron number.}
\label{tab:Neground}
\begin{center}
\begin{tabular}{c c|c c|c c c|c c c}
\hline \hline
$A$&$N$&$E$&$E/A$&$R_n$&$R_p$&$R_m$&$\beta_{2n}$&$\beta_{2p}$&$\beta_{2m}$\\
\hline
    18 & 8  & -135.584 & -7.532 & 2.555 & 2.951 & 2.782 & 0.000 & 0.035 & 0.020 \\
    19 & 9  & -144.546 & -7.608 & 2.686 & 2.840 & 2.768 & 0.312 & 0.434 & 0.377 \\
    20 & 10 & -157.932 & -7.897 & 2.805 & 2.839 & 2.822 & 0.518 & 0.533 & 0.526 \\
    21 & 11 & -166.982 & -7.952 & 2.867 & 2.823 & 2.846 & 0.513 & 0.507 & 0.510 \\
    22 & 12 & -176.806 & -8.037 & 2.928 & 2.813 & 2.876 & 0.508 & 0.487 & 0.499 \\
    23 & 13 & -182.704 & -7.944 & 2.958 & 2.789 & 2.885 & 0.368 & 0.408 & 0.386 \\
    24 & 14 & -189.974 & -7.916 & 2.994 & 2.758 & 2.898 & -0.275 & -0.235 & -0.258 \\
    25 & 15 & -194.881 & -7.795 & 3.088 & 2.786 & 2.971 & 0.256 & 0.326 & 0.284 \\
    26 & 16 & -200.902 & -7.727 & 3.180 & 2.807 & 3.042 & 0.277 & 0.327 & 0.296 \\
    27 & 17 & -204.405 & -7.570 & 3.253 & 2.821 & 3.100 & 0.224 & 0.306 & 0.254 \\
    28 & 18 & -209.152 & -7.470 & 3.333 & 2.836 & 3.165 & 0.191 & 0.289 & 0.226 \\
    29 & 19 & -211.688 & -7.300 & 3.376 & 2.850 & 3.205 & 0.117 & 0.234 & 0.157 \\
    30 & 20 & -215.708 & -7.190 & 3.428 & 2.863 & 3.251 & 0.055 & 0.160 & 0.090 \\
\hline
\end{tabular}%
\end{center}
\end{table}

\FloatBarrier
\begin{figure}[!ht]
\centering \includegraphics[height=16.0cm]{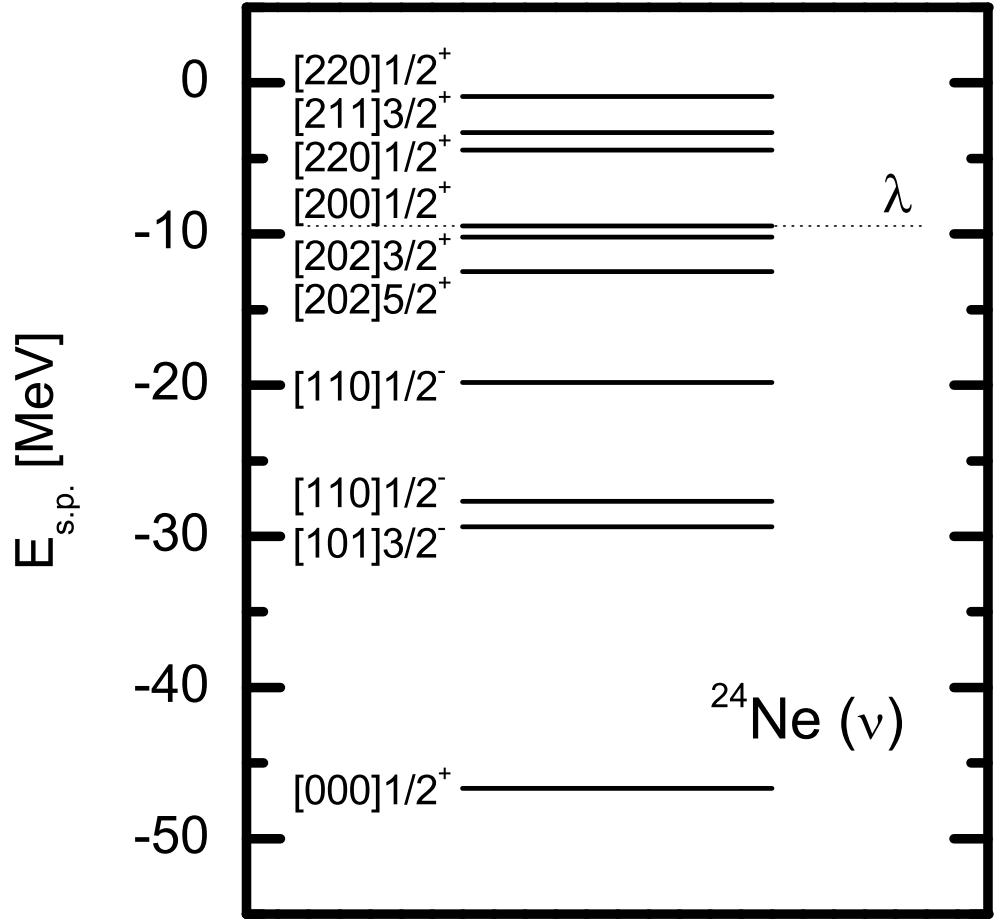} \vspace{12pt}
\caption{The neutron single-particle levels in $^{24}$Ne in RMF with
PK1. The dashed line represents the Fermi surface $\lambda$, which
is also the last occupied orbit in the simple filling
approximation.}
\end{figure}

\begin{figure}[!ht]
\centering \includegraphics[height=8.0cm]{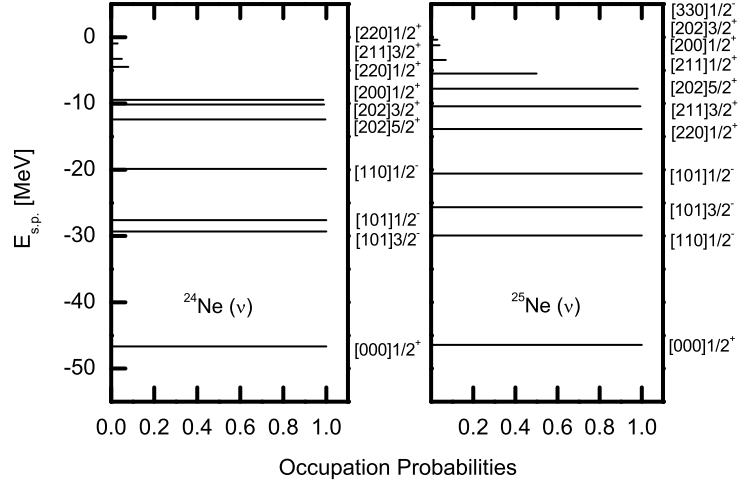} \vspace{0pt}
\caption{The occupation probabilities for the neutron
single-particle levels in $^{24}$Ne and $^{25}$Ne obtained from
RMF+SLAP calculation with PK1, where the occupation probabilities of
levels above [202]5/2$^+$ for $^{24}$Ne and those of levels above
[200]1/2$^+$ for $^{25}$Ne are multiplied by a factor of 10 for
clarity.}
\end{figure}

\begin{figure}[!ht]
\centering \includegraphics[height=8.0cm]{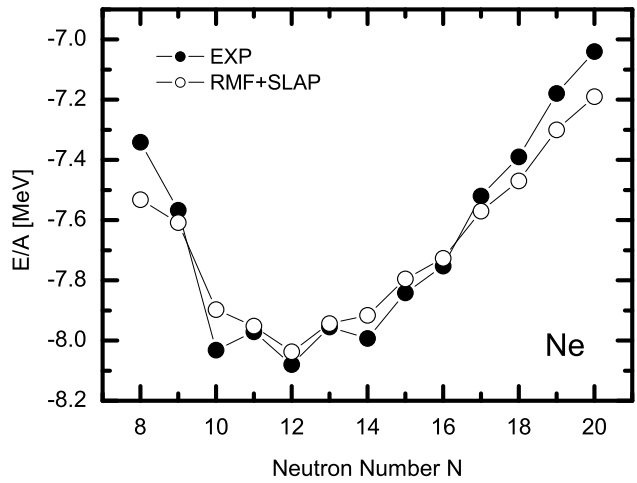} \vspace{12pt}
\caption{Binding energy per nucleon, $E/A$, for Ne isotopes as
functions of neutron number $N$ obtained from the RMF+SLAP
calculations (filled circles) with PK1 in comparison with the data
(open circles).}
\end{figure}

\begin{figure}[!ht]
\centering \includegraphics[height=8.0cm]{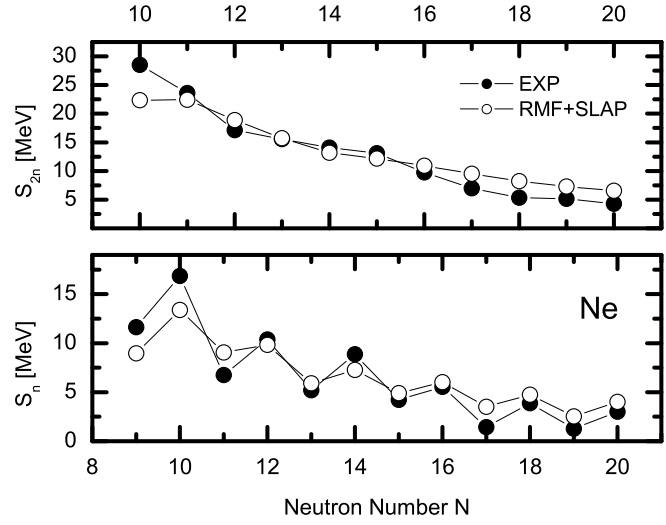} \vspace{12pt}
\caption{One- and two- neutron separation energies, $S_n$ and
$S_{2n}$, for Ne isotopes as functions of neutron number $N$
obtained from the RMF+SLAP calculations (filled circles) with PK1 in
comparison with the data (open circles).}
\end{figure}

\begin{figure}[!ht]
\centering \includegraphics[height=8.0cm]{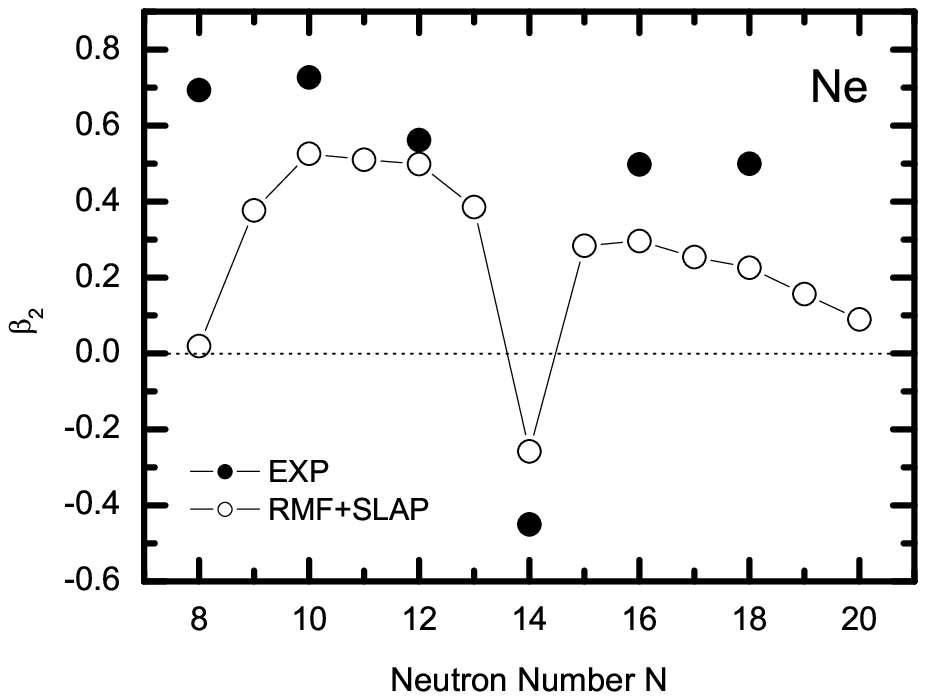} \vspace{12pt}
\caption{Quadruple deformation, $\beta_{2}$, for Ne isotopes as
functions of neutron number $N$ obtained from the RMF+SLAP
calculations (filled circles) with PK1 in comparison with the data
(open circles).}
\end{figure}

\begin{figure}[!ht]
\centering \includegraphics[height=8.0cm]{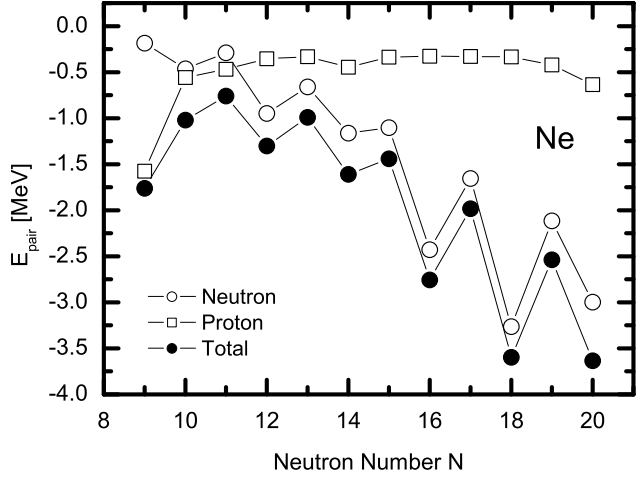} 
\caption{Pairing energies, $E_{pair}$, obtained from the RMF+SLAP
calculations with PK1 for Ne isotopes as functions of neutron number
$N$ for neutron (open circles), proton (open squares) and total
nuclear matter (filled circles).}
\end{figure}

\begin{figure}[!ht]
\centering
\includegraphics[height=10.0cm]{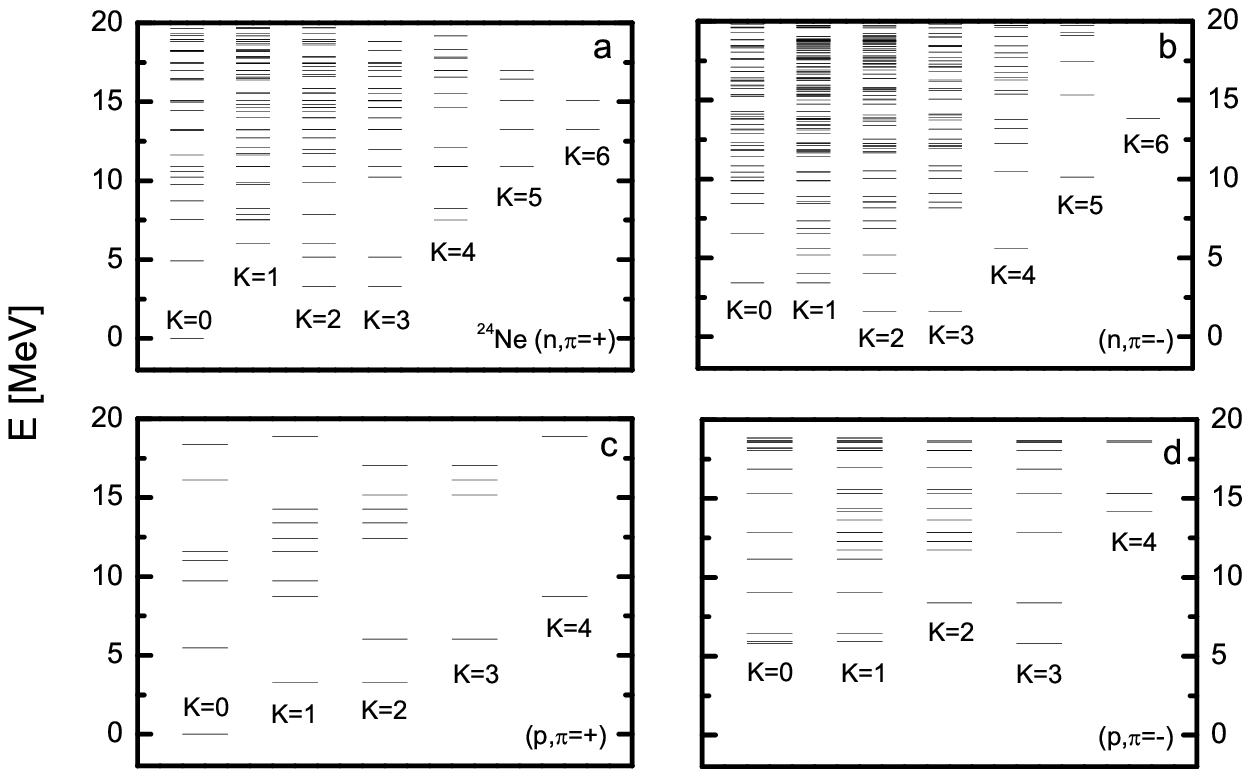}
\caption{The excitation spectra obtained from the
RMF+SLAP calculations with TMA in $^{24}$Ne. The quantum number $K$
is the eigenvalue for the third component of the total angular
momentum operator $J_z$. The subfigures (a), (b), (c) and (d)
respectively represent the spectra of positive and negative parity
for neutron and proton.}
\end{figure}


\begin{thebibliography}{99}

\bibitem{Serot86} B. D. Serot and J. D. Walecka, Adv. Nucl. Phys.
\textbf{16}, 1(1986).

\bibitem{Reinhard89} P. -G. Reinhard, Rep. Prog. Phys. \textbf{52},
439(1989).

\bibitem{Ring96} P. Ring, Prog. Part. Nucl. Phys. \textbf{37},
193(1996).

\bibitem{meng98npa} J. Meng, Nucl. Phys. A \textbf{635}, 3(1998).

\bibitem{meng05ppnp} J. Meng, H. Toki, S. G. Zhou, S. Q. Zhang, W. H. Long, L. S. Geng,
Prog. Part. Nucl. Phys. (2005) in press


\bibitem{Brockmann90} R. Brockmann and R. Machleidt, Phys. Rev. C \textbf{42},
1965(1990).

\bibitem{konig93} J. K\"onig and P. Ring, Phys. Rev. Lett. \textbf{71}, 3079(1993).

\bibitem{meng96prl} J. Meng and P. Ring, Phys. Rev. Lett. \textbf{77}%
, 3963(1996).

\bibitem{meng98prl} J. Meng and P. Ring, Phys. Rev. Lett. \textbf{80},
460(1998).

\bibitem{Arima69} A. Arima, M. Harvey, and K. Shimizu, Phys. Lett.
B \textbf{30}, 517(1969).

\bibitem{Hecht69} K. T. Hecht and A. Adler, Nucl. Phys. A
\textbf{137}, 129(1969).

\bibitem{Ginocchio97} J. N. Ginocchio,
Phys. Rev. Lett. \textbf{78}, 436(1997).

\bibitem{meng98prc} J. Meng, K. Sugawara-Tanabe, S. Yamaji, P.
Ring, and A. Arima, Phys. Rev. C \textbf{58}, R628 (1998).

\bibitem{meng99prc} J. Meng, K. Sugawara-Tanabe, S. Yamaji, and A.
Arima, Phys. Rev. C \textbf{59}, 154 (1999).

\bibitem{Zhou03prl} S. G. Zhou, J. Meng, and P. Ring, Phys. Rev.
Lett. \textbf{91}, 262501 (2003).

\bibitem{Mad00}H. Madokoro, J. Meng, M. Matsuzaki,
S.Yamaji, Phys.Rev. C \textbf{62}, 061301(2000).

\bibitem{Ma02}Z. Y. Ma, A. Wandelt, N. V. Giai, D. Vretenar,
P. Ring, L. G. Cao, Nucl. Phys. A \textbf{703} 222(2002).

\bibitem{Gambhir90} Y. K. Gambhir, P. Ring, and A. Thimet, Ann.
Phys. (N.Y.)\textbf{198}, 132(1990).

\bibitem{Geng03ptp} L. S. Geng, H. Toki, S. Sugimoto, and J. Meng,
Prog. Theor. Phys. \textbf{110}, 921(2003).

\bibitem{Dobaczewski84} J. Dobaczewski, H. Flocard, and J.
Treiner, Nucl. Phys. A \textbf{422}, 103(1984).

\bibitem{Dobaczewski96} J. Dobaczewski, W. Nazarewicz, T. R.
Werner, et al., Phys. Rev. C \textbf{53}, 2809(1996).


\bibitem{Yang01} S. C. Yang, J. Meng, and S. G. Zhou, Chin. Phys. Lett. \textbf{18}, 196(2001).

\bibitem{Cao02} L. G. Cao and Z. Y. Ma, Phys. Rev. C {\bf 66}, 024311 (2002).

\bibitem{Sandulescu03} N. Sandulescu, L. S. Geng, H. Toki, and G.
Hillhouse, Phys. Rev. C \textbf{68}, 054323(2003).

\bibitem{Zhang04} S. S. Zhang, J. Meng, S. G. Zhou, and G. C. Hillhouse
, Phys. Rev. C \textbf{70}, 034308(2004).

\bibitem{PY.57} R.E. Peierls and J. Yoccoz, Proc. Phys. Soc. {\bf A70}, 381 (1957)

\bibitem{RS.80} P. Ring and P. Schuck, {\em The Nuclear Many-Body Problem\/} (Springer Verlag, New York, 1980).

\bibitem{Zeh.67} H.D. Zeh, Z. Phys. {\bf 202}, 38 (1967)

\bibitem{HS95} K. Hara and Y. Sun, Int. J. Mod. Phys. {\bf E4}, 637(1995).



\bibitem{ER.82a} J.L. Egido and P.Ring, Nucl.Phys. {\bf A383}, 189 (1982)

\bibitem{ER.82b} J.L. Egido and P.Ring, Nucl.Phys. {\bf A388}, 19 (1982)



\bibitem{SR.00} J.A. Sheikh and P. Ring, Nucl. Phys. {\bf A665}, 71 (2000)

\bibitem{SR.02} J.A. Sheikh, P. Ring, E. Lopes and R. Rossignoli
Phys. Rev. C \textbf{66}, 044318 (2002).



\bibitem{Zeng83} J. Y. Zeng and T. S. Cheng, Nucl. Phys. A \textbf{405}%
, 1(1983).

\bibitem{Zeng941} J. Y. Zeng, Y. A. Lei, T. H. Jin, and Z. J. Zhao, Phys.
Rev. C \textbf{50}, 746(1994).

\bibitem{Rowe70} D. J. Rowe, $Nuclear$ $Collective$ $Motion$
(Methuen, London, 1970).

\bibitem{Chasman90} R. R. Chasman, Phys. Lett. B \textbf{242},
317(1990).

\bibitem{Molique97} H. Molique and J. Dudek, Phys. Rev. C \textbf{56}%
, 1795(1997).

\bibitem{Volya01} A. Volya, B. A. Brown, V. Zelevinsky, Phys. Lett. B \textbf{509},
37(2001)

\bibitem{Wu891prc} C. S. Wu and J. Y. Zeng, Phys. Rev. C \textbf{39}, 666(1989).

\bibitem{Wu91prl} C. S. Wu and J. Y. Zeng, Phys. Rev. Lett.
\textbf{66}, 1022(1991).

\bibitem{Zeng942} J. Y. Zeng, T. H. Jin, and Z. J. Zhao, Phys. Rev.
C\textbf{50}, 1388(1994).

\bibitem{Zeng01prc} J. Y. Zeng, S. X. Liu, Y. A. Lei, and L. Yu, Phys. Rev. C \textbf{63}, 024305(2001).

\bibitem{Zeng02prc} J. Y. Zeng, S. X. Liu, L. X. Gong, and H. B. Zhu, Phys. Rev.C \textbf{65}, 044307(2002).

\bibitem{Zhou03prc} S. G. Zhou, J. Meng, and P. Ring, Phys. Rev.
C \textbf{68}, 034323 (2003).

\bibitem{Long04} W. H. Long, J. Meng, N. V. Giai, and S. G. Zhou,
Phys. Rev. C \textbf{69}, 034319 (2004).

\bibitem{Sharma93} M. Sharma, M. Nagarajan, and P. Ring, Phys. Lett. B \textbf{312},
377(1993).

\bibitem{Lalazissis97} G. A. Lalazissis, J. K\"onig, and P.
Ring, Phys. Rev. C \textbf{55}, 540(1997).

\bibitem{Sugahara95} Y. Sugahara, Doctor thesi, Tokyo Metropolitan
University, 1995.

\bibitem{Geng04npa} L. S. Geng, H. Toki, A. Ozawa, and J. Meng,
Nucl. Phys. A \textbf{730}, 80(2004).

\bibitem{DelEstal01} M. Del Estal, M. Centelles, X. Vi\~{n}as,
and S. K. Patra, Phys. Rev. C \textbf{63}, 044321(2001).

\bibitem{Audi95} G. Audi and A. H. Wapstra, Nucl. Phys. A \textbf{595},
409(1995).

\bibitem{Raman01} S. Raman, C. W. Nestor Jr, P. Tikkanen, At. Data Nucl.
Data Tables \textbf{78}, 1(2001).

\bibitem{Zhang02} S. Q. Zhang, J. Meng, S. G. Zhou, and J. Y. Zeng,
Eur. Phys. J. A {\textbf 13}, 285(2002), and references therein.

\bibitem{mengplb98} J. Meng, I. Tanihata, S. Yamaji, Phys. Lett. B \textbf{419}, 1(1998).

\bibitem{mengprc02} J. Meng, H. Toki, J. Y. Zeng, S. Q. Zhang, and S.-G. Zhou,
Phys. Rev. C\textbf{65},  041302 (2002).

\bibitem{Baldwin97} G.
C. Baldwin and J. C. Solem, Rev. Mod. Phys. \textbf{69},
1085(1997).

\bibitem{Walker99} P. M. Walker and G. D. Dracoulis, Nature,
\textbf{399}, 35(1999).

\end{thebibliography}
\end{document}